\documentclass{elsart}

\usepackage{graphicx}
\usepackage{amssymb}

\begin{document}

\begin{frontmatter}

\title{Distribution of infected mass in disease spreading in scale-free networks}
\author{Lazaros K. Gallos and Panos Argyrakis}
\address{Department of Physics, University of Thessaloniki, 54124 Thessaloniki, Greece}

\date{\today}% It is always \today, today,
             %  but any date may be explicitly specified

\begin{abstract}
We use scale-free networks to study properties of the infected mass $M$
of the network during a spreading process as a function of the infection
probability $q$ and the structural scaling exponent $\gamma$. We use the
standard SIR model and investigate in detail the distribution of $M$, 
We find that for dense networks this function is bimodal, while for sparse 
networks it is a smoothly decreasing function, with the distinction between
the two being a function of $q$. We thus recover the full 
crossover transition from one case to the other. This has a result that 
on the same network a disease may die out immediately or persist for a 
considerable time, depending on the initial point where it was originated. 
Thus, we show that the disease evolution is significantly influenced by 
the structure of the underlying population.
\end{abstract}

\end{frontmatter}

There has been a growing interest recently in the network 
structure \cite{1,2,DM,3,4,5,6,7} and dynamics \cite{8,9} of real-life
organized systems. Many such systems, covering an extremely wide range 
of applications, have been recently shown \cite{1,2,DM,3,7} to exhibit 
scale-free character in their connectivity distribution, meaning that 
they obey a simple power law. Thus, the distribution of the connectivity 
of nodes, follows a law of the form
\begin{equation}
P(k) \sim k^{-\gamma} ,
\end{equation}
where $k$ is the number of connections that a node has and $\gamma$ is a 
network parameter which determines the degree of its connectivity.
These networks have some unusual properties, thus
justifying the heavy interest in recent years. For example,
spreading processes in scale-free networks show a dynamic behavior 
which is different than in other classes of networks. These processes 
are plausible models for the spreading of diseases, epidemics, etc.
Several models for spreading exist in the literature 
built on different algorithms, such as the SIR model \cite{Newman}, the SIS 
model \cite{8}, the SIRS model \cite{Kuperman}, etc. 

In a recent paper Newman \cite{Newman} studied analytically the SIR 
model in scale-free networks. The connectivity distribution had an 
exponential cutoff of the form
\begin{equation}
P(k) \sim k^{-\gamma} e^{-k/\kappa} \;,
\end{equation}
where $\kappa$ is an arbitrary cutoff value for $k$. This work gave
a closed form solution for the epidemic size and the average outbreak
size as a function of the infection probability. It showed that there 
is a critical infection threshold ($q_c$) only for small finite values 
of $\kappa$, but as $\kappa$ increases $q_c$ decreases, apparently 
resulting in $q_c\rightarrow 0$ (no critical threshold at all) in the limit of 
$\kappa \rightarrow \infty$. The same result was also shown for the SIS model by
Pastor-Satorras and Vespignani \cite{8}.

The absence of a critical threshold is not a universal network property.
Actually, in the well studied cases of lattice networks and small-world
networks the opposite is true \cite{Warren}. Such a threshold \cite{11,12} is
always present, which separates the infected from the uninfected regions. 
This has as a result that these models do not offer a very realistic picture. 
Recently, Warren {\it et al.} \cite{Warren} have used a heterogeneous 
distribution for the infection probabilities both for lattices and 
small-world networks. They model the variability in a population which 
results in a  broadening of the transition regime; however, a threshold
still exists and the behavior of the transition is qualitatively similar 
to the case of the simpler SIR model on a lattice. Because of this, 
scale-free networks are distinctly different regarding the predictions 
on the rate and efficiency of spreading. This is clearly much closer to
what it is intuitively expected, and can provide useful estimates for the 
properties of epidemics of any kind. The same type of model could also 
describe a diverse set of networks, such as social networks, virus spreading 
on the Web, rumor spreading, signal transmission etc.

In the present study we calculate the distribution of the epidemic size, 
i.e. the distribution of the infected mass, for several different $\gamma$ values. 
This property helps us to better understand the importance of the 
starting point (origin) of the disease. It turns out that this distribution
is not a smooth function for all networks, but depends strongly on the 
network density, i.e. the value of $\gamma$. 

We use a simulation algorithm to construct a scale-free network comprised 
of $N$ nodes. We follow a network generation method which enables us to 
freely vary the connectivity distribution of the network. We assign a number 
of edges $k$ for each node by using a power-law distribution $P(k)\sim k^{-\gamma}$. 
Starting from the highest connectivity nodes we create links by randomly 
choosing $k$ other nodes. Care is taken that no duplicate links are 
established between the same two nodes and once a node has reached 
the number of edges initially assigned to it, it no longer accepts 
any new connections. The cutoff value for the maximum possible 
connectivity of a node was fixed to $N/2$.

The spreading of a disease follows the standard SIR (Susceptible, 
Infective, Recovered) model. Initially, all nodes are in the susceptible
(S) status, and a random node is infected (I). During the first time 
step it tries to infect with probability $q$ the nodes linked to it,
and when the attempt is successful the status of the linked node 
switches from S to I. The process is repeated with all infected nodes 
trying to influence their susceptible (S) neighbors during each time step. 
After trying to infect its neighbors the status of an infected site 
changes to recovered (R) and can no longer be infected. The simulation 
stops when there are no infected nodes in the system or when all nodes 
have been infected. Small-world networks were constructed as 
described in reference \cite{10}.

\begin{figure}
\begin{center}
\includegraphics{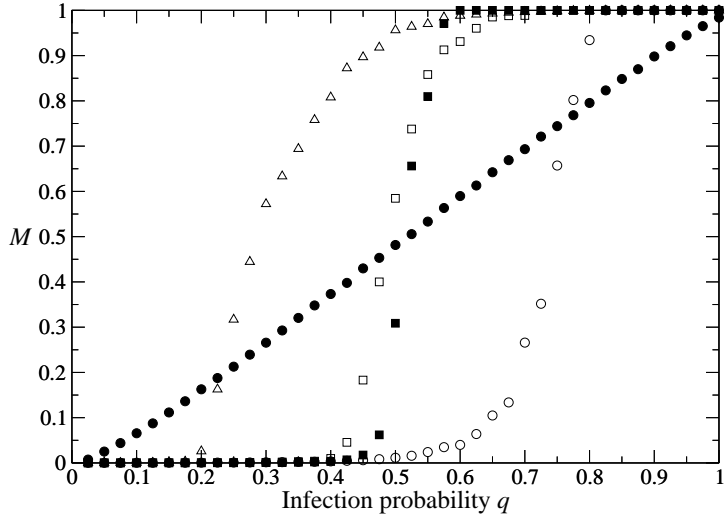}
\caption{Percentage $M$ of infected sites as a function of $q$,
for the different models studied ($\bullet$: scale-free network, $\gamma=2.0$, filled square: two-dimensional lattice,
$\bigcirc$: small-world $p$=0.0, $\Box$: small-world $p$=0.01, $\bigtriangleup$: small-world $p$=1.0, where $p$ is
the probability for rewiring a link in a small-world network).
The absence of a threshold is evident for scale-free networks.}
\end{center}
\end{figure}

We monitor the percentage $M$ of nodes infected and the duration 
of a disease (i.e. the time needed for the disease to either disappear 
or cover the entire network). In figure 1 we show for comparison purposes the percentage $M$ 
of infected sites as a function of $q$, for three different network types.
The form of the curves for the lattice and the small world 
networks is similar,
i.e. in all cases there is a sharp transition and a critical point.
However, this behavior is unrealistic, 
as it does not follow the majority of actual situations, such as
disease spreading in real-life networks \cite{Warren}. If this were to happen, e.g. the Web 
would be in a state of
either no virus present or the entire Web (all computers in the world) 
would be infected, with a very small probability of having an intermediate
situation with only a certain fraction of computers infected, which is 
the realistic picture.
Scale-free networks with a high degree of 
connectivity ($\gamma=2.0$) follow a much smoother spreading
evolution, as the mass of the infected population 
increases almost linearly with the infection probability q and there is no transition regime.
This is in agreement with the recent formalism of Newman \cite{Newman}.
The linear behavior can be understood as follows: On a scale-free network there
exist nodes with a wide range of connectivity. For fixed infection
probability $q$, the average probability for an infected node with $k$ links to
spread the disease is $kq$. If this number is greater than 1,
it is statistically certain that a neighbor node will be infected.
If it is significantly less than 1 the disease will die out. On the other hand, for a small-world network and
for a lattice network there is a characteristic mean number of links $\langle k \rangle$ assigned
to each node.

Here, for a scale-free network, we look at the distribution of the quantity $M$.
This is shown in figure 2, where we see that for different $\gamma$ values we have two opposite situations.
First, for $\gamma=2.0$ we see that for fixed $q$ we can have both cases coexisting, depending on the
connectivity of the initially infected node. For this case the distribution $\Phi(M)$ of the infected
population for fixed $q$ is bimodal and comprises of two distinct parts,
a strong peak at $M=1/N$ (only one node is infected) attributed to initially infected nodes with $k$=1,
or more accurately $kq\ll 1$, and a Gaussian-shaped part at some higher value of $M$.
The almost linear increase of $M$ with $q$ in figure 1 is due to the fact that for higher $q$ the peak of
$1/N$ decreases ($kq$ increases), and the average value of the Gaussian distribution increases accordingly.
We also notice in figure 2 that the gaussian part of the distribution has an average value which is larger
than $q$. For example, for $q=0.1$ the peak of the distribution is close to 0.25, and when $q=0.8$ it is
closer to 0.9. This means that if the peak in $1/N$ did not exist, the behavior of the curve
in figure 1 would be superlinear, i.e. $M$ would always be larger than $q$ as a result of the complex
connectivity of the network, but it would not reach the value of $M=1$ for infection probabilities less than 1.
The interplay between the peak at $1/N$ and the gaussian part yields the final curve which follows
roughly a linear increase.

\begin{figure}
\begin{center}
\includegraphics{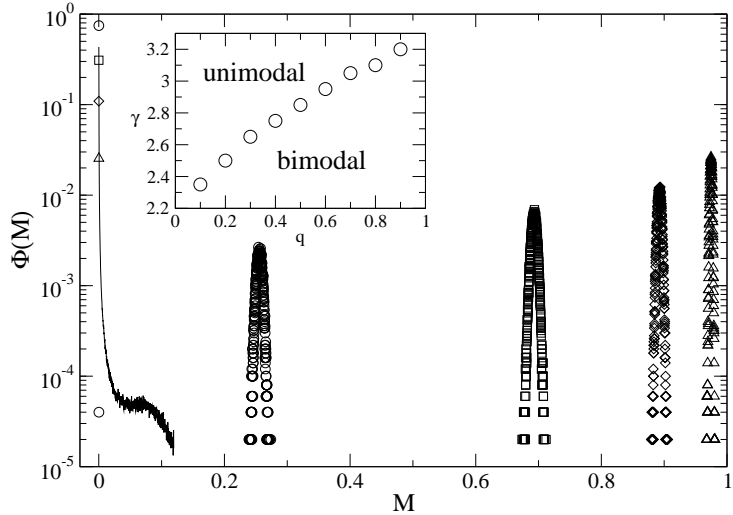}
\caption{Infected mass distribution $\Phi(M)$ of scale-free networks for $\gamma=2.0$, and different infection
probabilities: $q$=0.1 ($\bigcirc$), 0.5 ($\Box$), 0.8 ($\diamond$), 0.95 ($\bigtriangleup$).
To each gaussian part corresponds a single point at a value higher than
$M=1/N$, except for the case of $q=0.1$ which has two points (one at $1/N$ and one at $2/N$).
Solid line: $\gamma=2.9$ and $q=0.5$. Inset: Approximate values of $\gamma$ where the distribution
turns from bimodal to unimodal, as a function of $q$.}
\end{center}
\end{figure}

This result is a superposition of the two possible states present in the case of a simple lattice, where below the threshold the distribution
is a simple peak at $M=1/N$, while above the threshold the distribution peaks around $M=1$.
On a scale-free network there exists a finite probability for the disease to be either eliminated immediately or
cover a considerable portion of the network. Thus, for small $\gamma$ values,
the prediction on the future of a disease largely depends on the place where it originates, because of the
existence of the gap. This bimodality implies that every disease which survives the initial step(s) spreads over a non-zero portion of the
network population. This is not the case for large $\gamma$ values, where
recently \cite{15}, results for the SIR model on a scale-free network (with $\gamma=3.0$) were presented. The distribution of $M$ did not show any
bimodality, similarly to our solid curve on the left of figure 2, which corresponds to $\gamma=2.9$ and $q=0.5$.
We see that the bimodality is now lost, and it is replaced by a smoothly decreasing function
with increasing $M$. This result is in agreement with the work of Moreno {\it et al.}
\cite{15}. Thus, the distribution changes from bimodal to unimodal as we go from
a dense network to a sparse network. The crossover from one to the other takes
place as it is given in the inset in figure 2, where we plot $\gamma$ vs $q$, i.e. each
($\gamma$, $q$) pair is exactly at the corresponding transition point.

These observations show that in scale-free networks nodes of high connectivity 
act as ``boosters'' to the disease spreading;
even if very few nodes remain infected, by the time a high connectivity node is infected it spreads the disease over a significant number of its neighbors,
even for small $q$. This fact stresses the importance of the `hubs', as it has also been observed in the past in studies of the static properties for such networks
\cite{5,13,14}. Similarly, the low-connectivity nodes may serve to isolate large clusters of the network. These
clusters are effectively screened by the disease via the presence of the low-connectivity node, especially in the case of high $\gamma$, where the network is loosely connected.

Similar conclusions can be drawn for the duration of a disease. For small-world networks and regular lattices the duration of a disease when epidemics
takes place is practically constant for infection probabilities greater than the threshold (figure 3). For scale-free networks, on the contrary, there
is a slow increase of the duration as a function of $q$. However, the duration is much smaller now, which implies that even
for considerable infection probabilities a disease cannot last for a long time, and a considerable portion of the network can be infected
in a practically small and constant time. The duration of the disease at $q=1$ is also a measure of the network `diameter', since it represents
the average number of links needed to cross, before reaching all the system nodes.

\begin{figure}
\begin{center}
\includegraphics{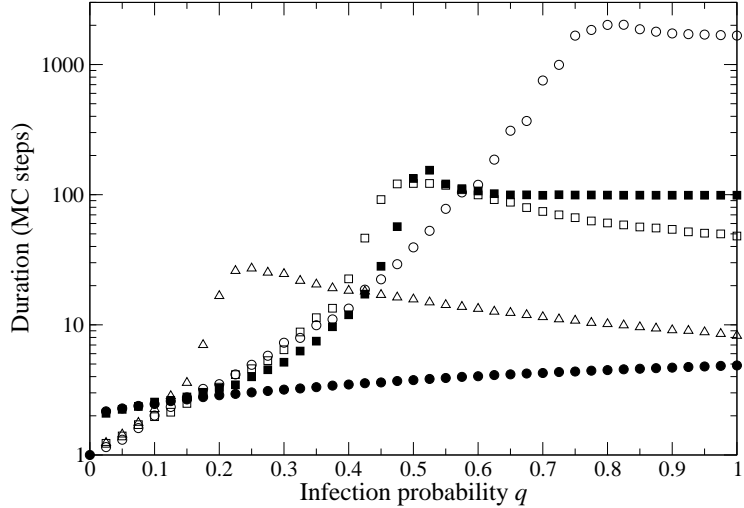}
\caption{Disease duration on different topologies. Symbols are the same as in figure 1.}
\end{center}
\end{figure}

Upon monitoring the distribution of uninfected
sites as a function of time we observed that it followed a power law at all times. The exponent of this power law was always the same as the one used for
the initial connectivity distribution, with the curve scaled down by a constant factor. Thus, sites of different connectivity are infected with the same relative rate.

Summarizing, we have investigated spreading properties on scale-free networks.
For the SIR model we studied we find that the starting point of the disease is 
very important, because it can either stop the disease or facilitate its spreading. 
This phenomenon for dense networks yields a bimodal distribution for the infected mass, 
with a peak close to 0 and a gaussian part around a finite value of $M$. 
Despite the smooth
increase of the infected mass with $q$ the disease spreads rapidly on
the network in a practically constant time, almost independently of $q$. 
This rapid spreading manifests the compactness of the
network (as compared to lattice and small-world networks) and its small 
diameter which is related to the short path length from
any site of the network to another. For sparse networks this is not the case, as it also
has been previously observed.

\end{document}